\documentclass{article}

\usepackage{epstopdf}% To incorporate .eps illustrations using PDFLaTeX, etc.
\usepackage{subfigure}% Support for small, `sub' figures and tables

\usepackage{natbib}% Citation support using natbib.sty
\bibpunct[, ]{(}{)}{;}{a}{}{,}% Citation support using natbib.sty
\renewcommand\bibfont{\fontsize{10}{12}\selectfont}% Bibliography support using natbib.sty

\usepackage{amsmath}
\usepackage{amssymb}
\usepackage{authblk}
\DeclareMathOperator*{\argmax}{argmax}
\usepackage{algorithm}
\usepackage{algpseudocode}

\usepackage{standalone}
\usepackage{wrapfig}
\usepackage{circuitikz}
\usepackage{tikz}
\usepackage{pgfplots}
\pgfplotsset{compat=1.18}

\begin{document}

\title{Neurons for Neutrons: A Transformer Model for Computation Load Estimation on Domain-Decomposed Neutron Transport Problems} %title of paper

\author[1]{Alexander Mote}
\author[2]{Todd Palmer}
\author[1]{Lizhong Chen}
\affil[1]{Oregon State University School of Electrical Engineering and Computer Science}
\affil[2]{Oregon State University School of Nuclear Science and Engineering}
%\name{Alexander Mote\textsuperscript{a}, Todd Palmer\textsuperscript{b}, and Lizhong Chen\textsuperscript{a}}
%\affil{\textsuperscript{a}Oregon State University School of Electrical Engineering and Computer Science, 2500 NW Monroe Ave, Corvallis OR 97331; \textsuperscript{b}Oregon State University School of Nuclear Science and Engineering, 1791 NW Campus Way, Corvallis OR 97331}
%}

\maketitle

\begin{abstract}
Domain decomposition is a technique used to reduce memory overhead on large neutron transport problems.
Currently, the optimal load-balanced processor allocation for these domains is typically determined through small-scale simulations of the problem, which can be time-consuming for researchers and must be repeated anytime a problem input is changed.
We propose a Transformer model with a unique 3D input embedding, and input representations designed for domain-decomposed neutron transport problems, which can predict the subdomain computation loads generated by small-scale simulations.
We demonstrate that such a model trained on domain-decomposed Small Modular Reactor (SMR) simulations achieves 98.2\% accuracy while being able to skip the small-scale simulation step entirely.
Tests of the model's robustness on variant fuel assemblies, other problem geometries, and changes in simulation parameters are also discussed.
\end{abstract}

%\begin{keywords}
%Monte Carlo; Neutron transport; Machine learning; Domain decomposition
%\end{keywords}

\section{Introduction}
\label{sec:intro}

Particle transport problems are a common application of high performance computing resources within the computational physics community.
GPU acceleration allows for these problems to be solved orders of magnitude more rapidly, particularly for Monte Carlo simulations, the operations of which are known to be ``embarrassingly parallel" \citep{ParComp}.
Traditional Monte Carlo simulation methods use a process known as domain replication, where each processor holds a copy of the full problem in memory, and tracks a subset of the particles within that problem; for GPUs, this leads to a memory bottleneck as the size and complexity of the problem increases.
This memory bottleneck is mitigated by the use of domain decomposition \citep{MCDD}, where each processor is instead given a portion of the full problem geometry (referred to as a ``subdomain") and only simulates particles within that subdomain.
Any particles that travel out of a processor's subdomain are communicated to the neighboring subdomain on a different processor using message passing.
This method of simulation allows GPUs to be efficiently utilized on large-scale neutron transport problems, such as \textit{k}-eigenvalue problems in modern nuclear reactors.

However, not all subdomains are created equal.
The composition of a neutron transport domain may create an imbalance in computation load across different subdomains; as such, uniformly assigning an equal number of processors to each subdomain in a given problem could drastically impact runtimes.
To balance this computational load, processors are instead allocated to subdomains such that each subdomain has a roughly equivalent overall runtime.
The computational load of each subdomain is typically estimated by way of a small-scale version of the simulation, which runs with some fraction of the total number of particles needed for the full problem.
However, this small-scale simulation can still be very computationally expensive and time-consuming for experimenters; additionally, new computational load estimates must be gathered when nearly any change is made to the input parameters of a simulation.
Even when the problem input itself is the same, hardware differences or increased particle interactions within a given subdomain can change the results between a small-scale and large-scale simulation.

This requirement to repeatedly generate computation load estimates demonstrates a need for a faster, but still equivalently accurate, method of estimating computation load for these large-scale problems.
Machine learning is known to efficiently generate rapid, accurate estimates once trained on a problem set; however, no model has been created or trained for particle transport problems, so it is unknown if machine learning is a valid method for this field of study.
Therefore, this paper sets out to create a ``proof-of-concept" machine learning model that can learn the patterns and relationships present in a particle transport domain, train it on a dataset of particle transport problems, and generate accurate estimations more quickly than the current state-of-the-art methodology.

To this end, we describe a deep learning model, specifically a novel Transformer model, which has been trained on domain-decomposed Small Modular Reactor (SMR) simulations, and predicts a computation load value for each subdomain in the full problem geometry.
This model is uniquely designed to handle 3-dimensional reactor domains, using a 3-dimensional input embedding and subdomain-based input representation, and utilizes the innate features of the Transformer model to learn the spatial relationships between the input elements and subdomains of a domain-decomposed particle transport problem.
These model's predictions represent an estimate of overall particle simulation time in each subdomain, which can then be used to generate processor allocations for large-scale simulations without needing to run small-scale versions first.

The results of this model are compared to current processor allocation methods on the Oak Ridge National Laboratory's (ORNL) Monte Carlo radiation transport code, Shift.
We show that the model's computational load predictions are generated orders of magnitude faster than the results of a small-scale Shift simulation, and have an overall accuracy of 98.5\% compared to the results of the small-scale simulations.
The results of an ablation study performed on the model, which allowed us to obtain the optimal hyperparameters for model training, are also discussed.
Tests of the model's robustness with respect to changes in fuel assemblies, fuel composition, and problem geometries show similar performance.
A preliminary study on predicting across simulations with multiple changes in parameters is also performed.
These studies and tests show that this model is a feasible replacement for the current computational load estimation method, and can be further developed to rapidly create accurate estimates for a number of particle transport problems and domains.

\section{Previous Work}
\label{sec:prev-work}

Shift \citep{Shift} is a state-of-the-art Monte Carlo radiation transport code designed at ORNL to simulate neutron physics.
It is massively parallel, and has demonstrated utilization of GPU architectures to achieve a 100-times speedup compared to CPU runtimes \citep{ShiftGPU}.
On larger problem domains, however, GPUs become infeasible, as the memory for all nuclides and tallies being simulated on a GPU must be allocated upfront.
To reduce this memory overhead, domain decomposition was implemented in Shift, allowing complex problem geometries, like that of a depleted SMR, to be simulated on GPUs \citep{ShiftDD}.

Monte Carlo code developers at ORNL have previously developed a processor allocation algorithm \citep{ProcAllo} for use with Shift.
This algorithm uses an ``allocation rank" diagnostic value that is derived from small-scale simulation data to estimate computation load; this value is a normalized estimation of particle simulation time.
An allocation rank value is generated for each subdomain during a simulation, and these values are used in a post-processing algorithm to balance computational load.
The pseudocode for this algorithm can be found in Algorithm \ref{alg:procallo}.

\begin{algorithm}
    \caption{Subdomain Processor Allocation}
    \begin{algorithmic}
    \State Set $N_{procs} = $ number of processors
    \State Set $N_d = $ number of subdomains
    \State Set $q = (1,1,\cdots,1)^T \in \mathbb{N}^{N_d}$
    \For{$k \in \{1, \cdots, N_{procs} - N_d\}$}
        \State Find $i^* = \argmax_{0 \leq i \leq N_d - 1}[(t_d)_i/q_i]$
        \State Increment $q_{i^*}$ by 1
    \EndFor
    \end{algorithmic}
    \label{alg:procallo}
\end{algorithm}

The algorithm works as follows: each subdomain is first assigned a single processor.
Then, the remaining processors are assigned one at a time to the subdomain with the current highest ratio of computational runtime ($t_d$) to number of allocated processors.
This allocation continues until all processors have been allocated to a subdomain; the number of processors allocated to each subdomain is then reported, and can be added to the large-scale simulation input file for correct allocation during the simulation proper.

For this paper, we consider the Nuscale SMR domain for our neutron transport problems; specifically, we are simulating a \textit{k}-eigenvalue problem using a ``fresh-fuel" version of the domain.
This domain was chosen as a benchmark for domain decomposition performance in previous papers \citep{ShiftDD}, and is relevant to the field of nuclear engineering as a state-of-the-art nuclear reactor that requires thorough large-scale simulations to assess its safety and optimize performance.
An existing simulation domain for this reactor was created by previous researchers \citep{ExaSMR}, using ORNL's VERA tool \citep{Vera} for simulating Light Water Reactors (LWRs).
The portion of the reactor domain we are concerned with simulating -- the core -- consists of an array of fuel assemblies.
These assemblies themselves consist of hundreds of fuel rods, or pins, as well as structural tubing material, arranged on a square grid.
Notably, these assemblies are all constructed on the same grid, and the type of assembly denotes the type of fuel used to fill the grid, usually specifying an amount of enriched uranium used within the fuel pins.
The structure of these assemblies informs the method we use to represent the problem geometry to the machine learning model in section \ref{ssec:domains}.

\section{Motivation}
\label{sec:motivation}

As discussed in section \ref{sec:intro}, domain decomposition requires a processor allocation algorithm to ensure that computational load is balanced across all subdomains in a simulation.
The processor allocation algorithm described in section \ref{sec:prev-work} achieves this by minimizing the proportion of subdomain load to processor allocation; in other words, a subdomain with a high computational load receives a greater proportion of the available processors than one with a low computational load.
During our initial survey, this algorithm was not identified as needing improvement; instead, we identify the small-scale simulations used to generate these computational loads as the slowest step in processor allocation, and the one that most needs improvement. This is because new computational load estimates must be gathered not only for new subdomain configurations or problem types, but for nearly any change to the input parameters of a simulation, leading to costly and time-consuming repetitions of this step.

As an example, Table \ref{table:params} shows the difference in small-scale computation load estimates of multiple simulations, all solving the same fresh-fuel SMR \textit{k}-eigenvalue problem on 16 subdomains, but with changes to the input parameters of the simulation.
These changes are shown as an average percent change compared to a control simulation with 400 MPI ranks, 20 cycles (10 inactive cycles followed by 10 active cycles), and one million particles.
This table also shows the percent change in runtime when the control simulation's computational load is used to allocate processors on the other simulations, rather than their own small-scale approximations; these full-scale simulations have the same parameters as those shown in the table, but with 50 times as many particle histories as their small-scale counterparts.
These differences demonstrate a need for a quick and computationally efficient method of approximating optimal processor allocations for subdomain load balancing, as the discrepancies between simulation results are significant enough that any change requires a new estimate of computational load.
It should be noted that the accuracy of these small-scale estimates is generally acceptable for practical applications \citep{ShiftDD}; it is the speed and efficiency of these approximations that need significant improvement.

\begin{table}[htbp]
\centering
\begin{tabular}{c|c|c}
\begin{tabular}[c]{@{}c@{}}Parameter\\ Change\end{tabular} & \begin{tabular}[c]{@{}c@{}}Average Change in\\ Processor Allocation\end{tabular} & \begin{tabular}[c]{@{}c@{}}Average Change in\\ Large-Scale Runtime\end{tabular} \\ \hline
800 MPI Ranks                                            & 9.33\%                                            & 11.33\%                                                                           \\
200 MPI Ranks                                           & 14.15\%                                            & 4.77\%                                                                           \\
30 Cycles                                                & 3.11\%                                            & 2.42\%                                                                           \\
10 Cycles                                               & 4.37\%                                            & 3.76\%                                                                           \\
2M Particles                                             & 6.24\%                                            & 6.67\%                                                                           \\
0.5M Particles                                            & 4.68\%                                            & 4.90\%                                                                          
\end{tabular}
\caption{Difference in processor allocations and large-scale runtimes when changing input parameters.}
\label{table:params}
\end{table}

Machine learning has been identified as a potential method to improve this speed and computational efficiency; machine learning algorithms are effective at detecting patterns and estimating target values when trained on a set of relevant input data points, and can do so rapidly once training is completed \citep{ML}.
Additionally, the small-scale simulations are generating these computational loads by simulating the same physical interactions and particle details as the large-scale simulation, which implicitly captures many of the details of the physics problem in their computation load output.
In other words, using these simulation outputs as targets for a machine learning model should allow it to generate accurate predictions without the need for detailed physical inputs.
However, the model does need to have an understanding of the physical space of the domain and the interactions between subdomains; because particles are communicated between subdomains as they move between boundaries, the model's representation of the problem geometry must be able to express communication overhead beyond the initial state of the problem.

Additionally, the model must be able to consider these spatial relationships across the entire domain, as there are limitations on where a particle can travel within a domain.
Many traditional deep learning models struggle with this; many models exist with short-term or local memory, but are not able to consider all inputs when making a prediction for a single element.
Due to this challenge, early attempts to use traditional artificial neural network (ANN) and convolutional neural network (CNN) architectures were unsuccessful, with prediction accuracy values below 75\% and simulation runtimes regularly ballooning to over double that of small-scale simulation results.
The Transformer model architecture \citep{Transformer} was identified due to its ability to perform calculations with respect to every element in an input, as well as its positional encoding method that can learn how these elements relate to one another spatially.
Transformers have proven success in the fields of computer vision and natural language processing due to these features; however, the reactor domains we are attempting to train with, and predict computational load for, contain some additional challenges that the Transformer model on its own cannot overcome.

First among these challenges is the variable number of subdomains found in the problem geometries used as inputs to this model.
Although some Transformer models have been developed to accommodate variable-length sentences, the three-dimensional domains and the arbitrary number of possible divisions of these domains into subdomains represent a far more complex input than a one-dimensional string of words.
Another challenge is representing this geometry to the model such that the model can actually learn from it.
There are several elements of the problem geometry that can be included to give the model insights into the problems we are trying to solve, but choosing the correct elements and organizing them in a way that is useful to the model is a challenge.
Ideally, these input elements would be able to represent a wide array of different reactors and problem types to the model.
Additionally, we must decide how much data is needed for the model to train successfully, keeping in mind that every point of data requires a small-scale simulation to retrieve target values for our training dataset.
The model itself must also be adjusted and optimized for this particular problem; no model can be taken ``off the shelf" for this application, as the specifics of this problem have not been addressed by any existing model.

Lastly, there is the issue of dataset generation and training time.
Although a trained model should be able to generate predictions faster than a small-scale simulation, this savings is greatly reduced if the model must be re-trained and new data must be generated for every problem that a given user would want to simulate.
For this reason, the model we create to replace the small-scale simulation must be able to be ``pre-trained"; this is a procedure in machine learning where a model uses a great deal of data and training time to learn the fundamental patterns of a given problem.
This model is then re-trained on a specific use case that requires only minor changes to the model's pre-trained weights, done via a shorter training run with a much smaller, case-specific dataset, in order to learn the more complex and unique patterns present in that use case.
Ideally, such a model could be pre-trained on a number of different problem domains, and then refined for the specific type of problem or even the exact domain the user will be simulating.
For the goals of this paper, however, the scope of this pre-training will be limited to a small selection of domains, with the model's robustness being tested by modifying these domains in ways not present in the model's training data; this is discussed further in section \ref{ssec:variants}.

\section{Methodology}
\label{sec:method}

\subsection{Representing the Domains}
\label{ssec:domains}

The NuScale SMR domain consists of a 7x7 grid of ``core assemblies", each containing different materials.
Some of these materials are nuclear fuel rods, which generate the particles we are concerned with simulating; these rods have varying amounts of enriched uranium that are defined by the user when the domain is created in VERA.
These assemblies extend uniformly in the z-axis of the reactor; therefore, the z-axis can be arbitrarily subdivided for presentation to our model without changing the material properties of these assemblies.

We define an elementary volume within the domain as the smallest volume that a domain can be subdivided into; in other words, the elementary volume is the ``pixel" of our input representation, defining the resolution with which our model can ``see" the domain.
For this problem, we take each core assembly and divide it along the z-axis into 7 equivalent pieces.
This one-seventh of a core assembly is our elementary volume for this problem, creating a 7x7x7 3-dimensional grid of elementary volumes that can be grouped into subdomains.

Transformer models accept ``embedded" input tokens of any value; this means that a reactor with complex fuel compositions or a multitude of nuclides that contribute to computational load could be represented using an embedding method similar to the one already used in VERA.
For this model, the enriched fuel concentration of each of these elementary volumes was selected as the embedding value for the model; volumes with no enriched fuel received a small non-zero value representing the particles that may travel through these volumes during the simulation.
These embedding values are then modified slightly based on their position along the z-axis; because these values are otherwise identical along the z-axis, this is done to communicate
further spatial information to the model during training.
Each volume's fuel concentration value is stored in a PyTorch tensor, to be later used to generate a model input once a domain decomposition has been generated.

\subsection{Dataset Generation}
\label{ssec:dataset}

A script was created that randomly generates a domain decomposition that aligns with the 7x7x7 grid created in section \ref{ssec:domains}; that is to say, each subdomain contains at least one elementary volume, and each elementary volume can be assigned to one and only one subdomain.
Random generation was used to ensure that the dataset represented a subset of the entire possible search space of subdomains, and reduced the risk of overfitting on selected or conventionally used decompositions.
This domain decomposition information is saved to a file that can later be used to run a Shift \textit{k}-eigenvalue simulation on this subdomain allocation; aside from this domain decomposition, the simulations are identical.

\begin{wrapfigure}{r}{0.5\textwidth}
    \centering
        \includegraphics[width=0.48\textwidth]{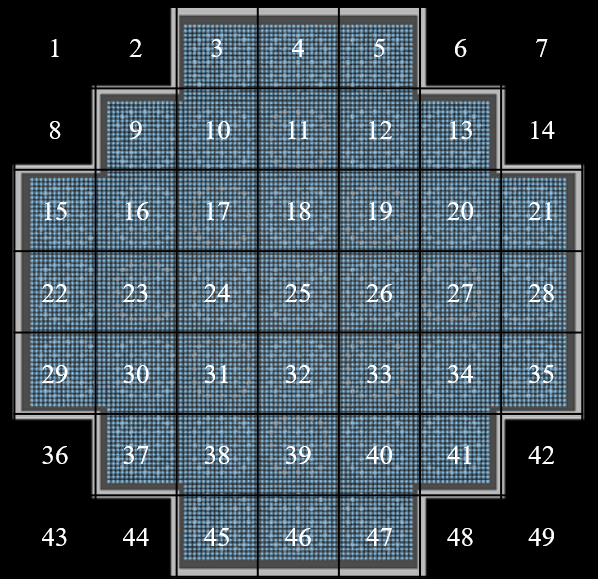}
    \caption{A z-axial slice of the SMR domain, and the indexing used for dataset generation.}
    \label{fig:smr}
\end{wrapfigure}

A new tensor is generated for this domain decomposition, containing a row for every subdomain in the simulation.
Each row is filled with the enriched fuel concentrations of all elementary volumes contained in that subdomain; these volumes are indexed using the method seen in Figure \ref{fig:smr}.
All other elements of the row are filled with a constant embedding value representing particle communication between ``out-of-subdomain" elements.
After the last subdomain is filled in this way, a row of ``end of sequence" values is added to inform the model that it has reached the final subdomain in this input.
Once filled with the relevant input information, these tensors are then padded with zeroes to ensure that they are all the same size for easier batch processing during training; this allows the transformer model to process inputs with an arbitrary number of subdomains.

Once the domain decompositions have been generated, a small-scale simulation is performed for every decomposition.
These simulations used 1 million particle histories, and were performed on 800 MPI ranks on Lawrence Livermore National Laboratory's Lassen machine; each simulation completed in approximately two minutes.
Each simulation executed with a uniform processor allocation; in other words, each subdomain received an equal number of processors, and any excess MPI ranks that could not be allocated in this way were unused.
The output files of these simulations contain an ``allocation rank" value for each subdomain in the simulation.
This allocation rank, used by Shift's current processor allocation algorithm to determine optimal load balancing, is a useful target for predicting computational load in these simulations, particularly for comparing the model's results to the current state-of-the-art methodology.
For this reason, these allocation rank values are collected after simulation and used as the output targets for our model.
After collection is complete, a machine learning dataset is traditionally normalized in some way to flatten large amounts of variance, remove outliers, and reduce training time.
However, the allocation rank values produced by one simulation do not directly correlate to the values produced by another, making normalization without information loss difficult; as such, these output targets are passed to the model as is, and neither the inputs nor targets are normalized.

It should be noted that all simulations for this paper were performed on CPU architecture; although domain decomposition does not provide as significant an improvement on CPUs as it does on GPU architectures, the computational load values obtained from these simulations have been shown to be comparable to those obtained from GPU simulations \citep{ShiftGPU}. 
Performing these simulations on CPU architecture allowed us to not only use a more stable version of Shift, but also run our simulations on a larger number of processors, leading to greater granularity for processor allocation.
This allowed for a more meaningful comparison of large-scale simulation runtimes in section \ref{sec:results}.

\subsection{Model Architecture}
\label{ssec:architecture}

\begin{wrapfigure}{r}{0.5\textwidth}
    \centering
    \resizebox{0.5\textwidth}{!}{%
    \begin{circuitikz}
        \tikzstyle{every node}=[font=\footnotesize]
        \draw [, dashed] (16.5,8.5) rectangle  (20.5,13.5);
        \draw [rounded corners = 15.0] (12.5,13.25) rectangle (16,12.25);
        \draw [rounded corners = 15.0] (12.5,11.5) rectangle (16,10.5);
        \draw [->, >=Stealth] (16,11) -- (16.75,11);
        \draw [rounded corners = 15.0] (16.75,13.25) rectangle (20.25,12.25);
        \draw [, dashed] (12.25,10.25) rectangle  (16.25,13.5);
        \draw [->, >=Stealth] (14.25,14) -- (14.25,13.25);
        \node [font=\footnotesize] at (16.35,14.5) {Positional Encoding};
        \draw  (13.5,16) rectangle (15,15);
        \draw  (17.75,16) rectangle (19.25,15);
        \node [font=\footnotesize] at (13.1,13.75) {Encoding (7x)};
        \node [font=\small] at (19.75,13.75) {Decoding (7x)};
        \draw (14.25,15) to[sinusoidal voltage source, sources/symbol/rotate=auto] (14.25,14);
        \draw (18.5,15) to[sinusoidal voltage source, sources/symbol/rotate=auto] (18.5,14);
        \draw [->, >=Stealth] (18.5,14) -- (18.5,13.25);
        \node [font=\footnotesize] at (18.5,12.95) {Masked};
        \node [font=\footnotesize] at (18.5,12.6) {Multi-Head Attention};
        \node [font=\footnotesize] at (14.25,12.75) {Multi-Head Attention};
        \node [font=\small] at (14.25,15.65) {Input};
        \node [font=\small] at (14.25,15.3) {Domains};
        \node [font=\small] at (18.5,15.7) {Output};
        \node [font=\small] at (18.5,15.3) {Targets};
        \draw [->, >=Stealth] (14.25,12.25) -- (14.25,11.5);
        \draw [->, >=Stealth] (18.5,12.25) -- (18.5,11.5);
        \node [font=\normalsize] at (14.25,11) {Linear Layer};
        \draw [rounded corners = 15.0] (16.75,11.5) rectangle (20.25,10.5);
        \draw [->, >=Stealth] (18.5,10.5) -- (18.5,9.75);
        \draw [rounded corners = 15.0] (16.75,9.75) rectangle (20.25,8.75);
        \node [font=\normalsize] at (18.5,9.25) {Linear Layer};
        \node [font=\footnotesize] at (18.5,11) {Multi-Head Attention};
        \draw [->, >=Stealth] (16.25,9.25) -- (16.25,8.25);
        \draw [rounded corners = 15.0] (14.5,8.25) rectangle (18,7.25);
        \node [font=\normalsize] at (16.25,7.75) {Linear Layer};
        \draw [](16.25,9.25) to[short] (16.75,9.25);
        \draw [->, >=Stealth] (16.25,7.25) -- (16.25,6.5);
        \draw [rounded corners = 15.0] (14.5,6.5) rectangle (18,5.5);
        \node [font=\normalsize] at (16.25,6) {PReLU};
        \draw [->, >=Stealth] (16.25,5.5) -- (16.25,4.75);
        \draw  (15.25,4.75) rectangle (17.25,3.75);
        \node [font=\small] at (16.25,4.45) {Model};
    \node [font=\small] at (16.25,4.05) {Predictions};
    \end{circuitikz}
    }%
    \caption{A diagram detailing the architecture and flow of the machine learning model.}
    \label{fig:model}
\end{wrapfigure}
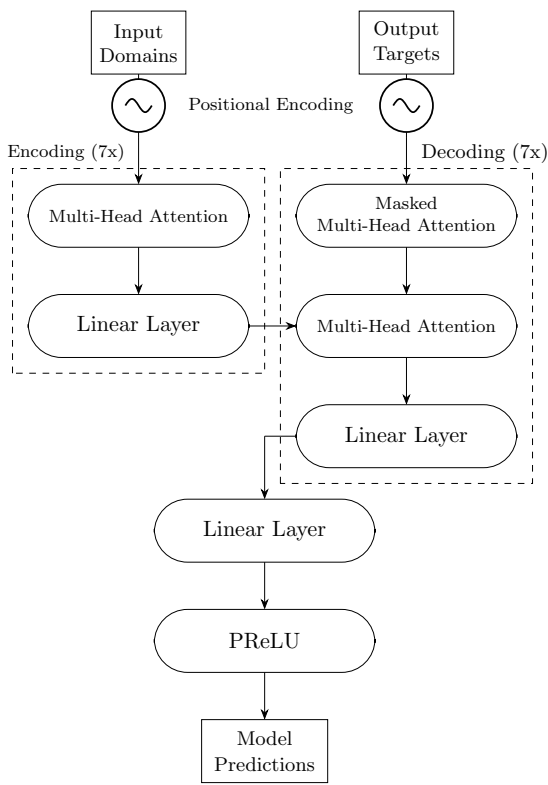

As mentioned in section \ref{sec:motivation}, a Transformer model was identified as the ideal machine learning model for this application.
The model used in this paper was constructed using PyTorch modules; Figure \ref{fig:model} shows the architecture and flow of the Transformer model.
Normally, Transformer models accept natural language sentences as an input, requiring the words to be ``embedded" into numeric tokens before being processed by the model; however, our inputs are already embedded into numeric values as described in section \ref{ssec:domains}, so this step can be skipped.
The first major feature of the Transformer model, positional encoding, is then applied.
The mathematical functions of this layer are shown in Equations \ref{eq:posenc1} and \ref{eq:posenc2}.
This feature slightly modifies the value of each input element based on their position in the input array.
In this case, it applies two sinusoidal functions to the array, creating a unique combination of positional values for every dimension in the array.
The positional values are incorporated into the model's calculations during encoding and decoding, allowing it to use the position of each input element to improve its predictions during training.

\begin{equation}
    PE_{pos, 2i} = \sin(pos/10000^{2i/343})
    \label{eq:posenc1}
\end{equation}
\begin{equation}
    PE_{pos, 2i+1} = \cos(pos/10000^{2i/343})
    \label{eq:posenc2}
\end{equation}

Sinusoidal functions are used to bound the position values within a cyclic function, allowing the model to accept inputs of arbitrary length; the number 10000 in the dividend is used to extend the wavelength of these sinusoidal functions and achieve unique positional values along all dimensions and all possible array input sizes.
Once positional encoding has been completed, the inputs are passed through seven encoding and decoding layers.
These layers utilize the second major feature of the Transformer model, known as multi-head attention.
The attention function is shown in Equation \ref{eq:attention}; much like human cognitive attention, it calculates the contextual relevance of every input element with respect to every other input element as a soft weight.
The soft weights are created by taking the product of every input element with every other input element; these products are referred to as a ``query-key (Q-K) pair".
The query-key pairs are then scaled by the length of the input array to minimize outliers, before a $softmax$ function is used to convert this value into a soft weight, effectively creating a percentage value representing the contextual relevance of each pair of input elements.
Finally, these soft weights are multiplied by a value ($V$) matrix, which further amplifies pairs with high relevance; this value represents the final attention calculation, and is the output of a single ``head" of our attention layer.

\begin{equation}
    Attention(Q,K,V) = softmax(\frac{QK^T}{\sqrt{d_k}})V
    \label{eq:attention}
\end{equation}

Each of these query, key, and value matrices also have their own corresponding weight matrix, which is adjusted during training to modify the values of this calculation to optimize the model's predictions.
Multi-head attention, shown in equation \ref{eq:multihead}, allows us to project the query, key, and value matrices multiple times, and adjust multiple weight matrices in parallel.
This projection reduces the dimensionality of each individual matrix, leading to a similar computational cost with the added benefit of parallel processing.
In this case, the model uses 7 heads for each multi-head attention layer; the results of each head are concatenated to form the full attention matrix after all heads are finished.

\begin{equation}
    MultiHead(Q,K,V) = Concat(head_1,\ldots,head_7)W^O
\label{eq:multihead}
\end{equation}
\begin{equation*}
    \text{where }head_i = Attention(QW_i^Q, KW_i^K, VW_i^V)
\end{equation*}

Each multi-head attention layer is followed by a simple linear layer, which applies a linear transformation to the attention values based on a weight matrix that is similarly adjusted during training.
Due to the multiple layers in each encoding and decoding block, these linear transformations are executed a total of 14 times whenever a prediction is generated by the model.
In the decoding block, the output targets are used during training, and the first multi-head attention layer is masked so that the model can only use the output values preceding the element it is currently predicting.
This allows the model to use accurate values when performing attention during training, without ``spoiling" the results that the model is attempting to predict.
When the model has finished training and is being evaluated, these output targets are replaced with the model's own predictions.

After the input is encoded and decoded by the Transformer model, a final linear layer is added to synthesize the decoded outputs for each subdomain into a single prediction for the computation load value of that subdomain.
These predictions are then passed through a ``parametric rectified linear unit" (PReLU) activation function to ensure that the values are largely positive.
The PReLU activation function, shown in Equation \ref{eq:prelu}, compresses negative predictions by some constant value based on a parameter $\alpha$ that is learned by the model during training, and allows negative values to contribute to weight calculations during training without skewing the final prediction values.

\begin{equation}
    PReLU(x) = \begin{cases}
     x, & x\geq 0 \\
     \alpha x, & o.w.
\end{cases}
\label{eq:prelu}
\end{equation}

After the model has been successfully trained, the average absolute error of the model's predictions on the training set is extracted, and this value is added to predictions made on the testing set.
This is motivated by an observation that subdomains with a small computation load are incredibly sensitive to this error; adding this error value to all predictions reduces this sensitivity on small-load subdomains without heavily impacting other subdomains, and improves results when the predictions are input into the processor allocation algorithm.

\section{Results}
\label{sec:results}

\subsection{Ablation Study}
\label{ssec:ablation}

\begin{figure}[htbp]
    \centering
    \input{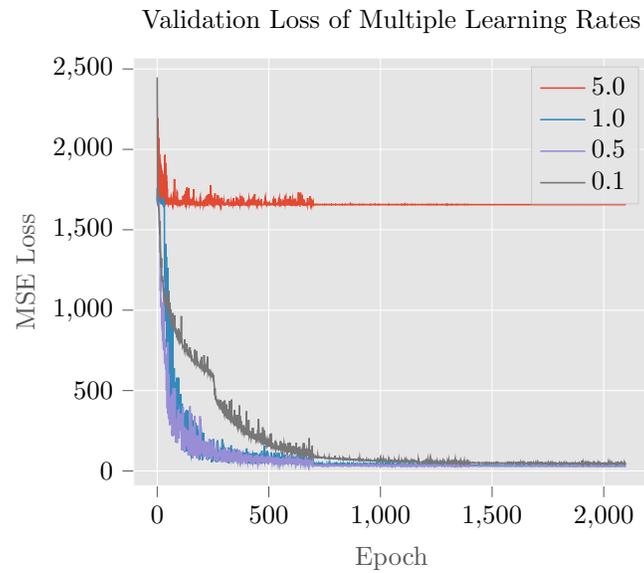}
    \caption{A comparison of validation losses across different initial learning rates.}
    \label{fig:lr}
\end{figure}

An ablation study \citep{ablation} was performed on the model to ensure that the hyper-parameters used for training were the optimal values for this application.
These studies are regularly used in machine learning to examine the effect of modifying each hyper-parameter within a given range and determine its impact on model performance.
All studies trained the model for 2100 epochs, stepping down the size of the learning rate every 700 epochs.
Figure \ref{fig:lr} shows the validation loss of the model across 2100 epochs when a number of different initial learning rates were used.
The minimum loss value from these candidates arose from a learning rate of $0.5$; this learning rate gave us a minimum loss value of $27.04$ using a Mean Squared Error (MSE) loss function.

\begin{figure}[htbp]
    \centering
    \input{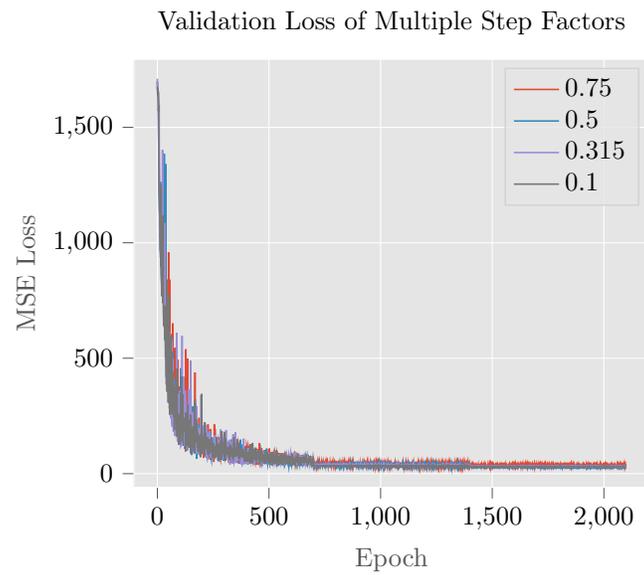}
    \caption{A comparison of validation losses across different learning rate step factors.}
    \label{fig:sf}
\end{figure}

Figure \ref{fig:sf} shows the validation loss of the model when a number of different learning rate step factors were used.
These step factors scale the learning rate as training continues, allowing the model to make large initial steps required for accuracy on large targets before focusing on smaller error corrections later in training.
These tests all used the same initial learning rate of $0.5$.
The minimum loss values were largely similar across all learning rates, with the step factor of $0.5$ slightly edging out the other candidates.
This step factor gives us a new minimum loss value of $25.48$.

\begin{figure}[htbp]
    \centering
    \input{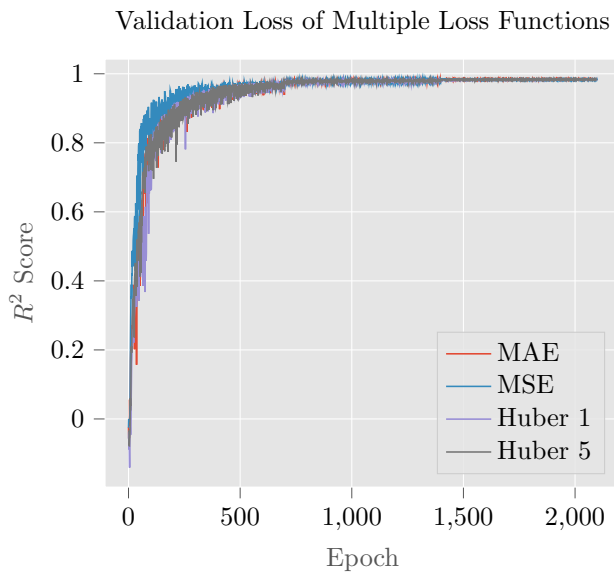}
    \caption{A comparison of $R^2$ scores across different loss functions.}
    \label{fig:func}
\end{figure}

Figure \ref{fig:func} shows the $R^2$ score of the model when a number of different loss functions were used.
The $R^2$ score is a common metric of goodness-of-fit in machine learning and statistical analysis, representing how well a model's prediction replicates its target values.
The loss functions chosen as candidates for this project were mean absolute error (MAE), mean squared error (MSE), Huber loss with a $\delta$ of 1.0, and Huber loss with a $\delta$ of 5.0.
MSE is a standard loss function in machine learning, but is known to be sensitive to outliers.
Huber loss is a function that combines MSE and MAE loss, using MSE when the absolute error is below its delta value, and MAE when the error is above it.
This is done to remove the outlier sensitivity found in MSE loss.
The MAE loss function is least sensitive to outliers, but has a ``sharper" gradient close to zero, which introduces some unwanted noise as error values get sufficiently small.
Because the loss values provided by these functions are not comparable, the $R^2$ score on the validation dataset was chosen as a metric for loss functions instead.
Much like the step factor study, there was little change in this value between candidates; MAE loss managed to provide the highest results with an $R^2$ score of $98.61\%$, but the other functions performed nearly equivalently, making the results difficult to compare in the figure.
The Mean Absolute Error loss function is seen in equation \ref{eq:mae}:

\begin{equation}
    MAE(x) = \sum^n\frac{|x_n - y_n|}{n}
    \label{eq:mae}
\end{equation}

where $x$ is the model's prediction, $y$ is the target value, and $n$ is the number of outputs.

\begin{figure}[htbp]
    \centering
    \input{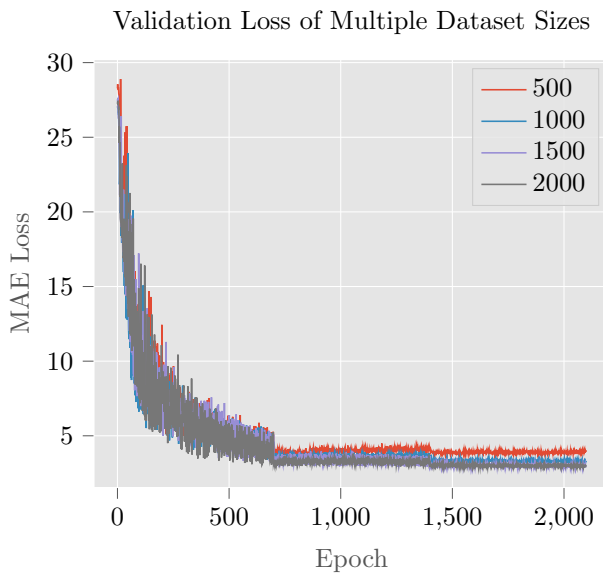}
    \caption{A comparison of loss values across different dataset sizes.}
    \label{fig:data}
\end{figure}

Lastly, Figure \ref{fig:data} shows the validation loss of the model when a number of different dataset sizes were used.
This test is designed to ensure that the model has sufficient data to reach its optimal predictions, but can also find a set of data points that may be more effective at training the model.
In this case, we can see that the loss values achieved with 1500 datapoints are similar to the loss values achieved with 2000 datapoints.
This means that simply increasing the number of points in the dataset globally would likely not significantly improve the model's predictions.
However, the minimum loss values of these two dataset sizes are almost identical, meaning the model is not learning more effectively with 1500 datapoints than it is with 2000.
This confirms that 2000 datapoints is a sufficient size for the model's dataset for the problem set we are analyzing; if this problem set was expanded to multiple domains or problem types, more data would likely be required.

\subsection{Fresh-Fuel SMR Results}

\subsubsection{Model Performance}
\label{ssec:performance}

\begin{figure}[htbp]
    \centering
    \input{model_loss}
    \caption{A graph of the model's training and validation losses during a training run on the fresh-fuel SMR dataset.}
    \label{fig:results}
\end{figure}

Figure \ref{fig:results} shows the training and validation losses of the model's training run over 3000 epochs on four AMD MI-250X GPUs; this training run lasted approximately six hours on this hardware.
The model started this run with a learning rate of 0.5, reducing by a factor of 0.5 every 1000 epochs.
This training run used a Mean Absolute Error (MAE) loss function, shown in equation \ref{eq:mae}.
The model achieved a minimum validation loss of 3.42; this loss was achieved on non-normalized output targets, with a range of values in the single-digits to hundreds, which leads to larger loss values than traditionally seen in machine learning applications.
The model's accuracy was also measured with an $R^2$ score; across the entire validation set, the model achieves an $R^2$ score of $98.23\%$.
The model was trained on 2000 datapoints; each point represents an input/output pair of an SMR domain decomposition and its corresponding computation load, as described in section \ref{ssec:dataset}.

One other area of performance to consider is the runtime of the model itself; the model must be able to generate predictions at the same speed as small-scale simulations or faster in order to be useful to researchers.
On average, the small-scale simulations used in the model's dataset took 112.5 seconds to complete, using 800 IBM Power9 CPU cores.
Comparatively, the Transformer model takes 55 seconds to generate predictions on four AMD MI-250X GPUs; however, the majority of this time is spent initializing the model and loading saved weights from training.
Once initialized, this model can generate predictions in under a second; during training, the model was regularly able to generate 200 predictions in 0.6 seconds, for an average of 2 milliseconds per prediction on four AMD MI-250X GPUs.
Even without considering the difference in computing power required for these two generation methods, this represents a speedup of several orders of magnitude over the existing methodology.

\subsubsection{Simulation Runtimes}
\label{ssec:runtimes}

The following results focus on three resolutions of domain decomposition that would realistically be used in a \textit{k}-eigenvalue simulation: 8, 16, and 32 subdomains.
Multiple simulations were performed for each of these resolutions, all with the subdomain divisions placed at different positions along the $x$-, $y$-, and $z$-axes.

\begin{figure}[ht]
    \centering
        % This file was created with tikzplotlib v0.10.1.
    \begin{tikzpicture}
    
    \definecolor{chocolate2267451}{RGB}{226,74,51}
    \definecolor{dimgray85}{RGB}{85,85,85}
    \definecolor{gainsboro229}{RGB}{229,229,229}
    \definecolor{lightgray204}{RGB}{204,204,204}
    \definecolor{steelblue52138189}{RGB}{52,138,189}
    
    \begin{axis}[
    axis background/.style={fill=gainsboro229},
    axis line style={white},
    legend cell align={left},
    legend style={
      fill opacity=0.8,
      draw opacity=1,
      text opacity=1,
      at={(0.03,0.03)},
      anchor=south west,
      draw=lightgray204,
      fill=gainsboro229
    },
    tick align=outside,
    tick pos=left,
    title={8-Subdomain Runtime Comparisons},
    x grid style={white},
    xmajorgrids,
    xmin=-0.283, xmax=5.943,
    xtick style={color=dimgray85},
    xtick={0.33,1.33,2.33,3.33,4.33,5.33},
    xticklabels={\footnotesize 2x2x2,\footnotesize 4x2x1,\footnotesize 2x4x1,\footnotesize 1x4x2,\footnotesize 4x1x2,\footnotesize Average},
    y grid style={white},
    ylabel=\textcolor{dimgray85}{Average Cycle Time (seconds)},
    ymajorgrids,
    ymin=50, ymax=110,
    ytick style={color=dimgray85}
    ]
    \draw[draw=none,fill=chocolate2267451,very thin] (axis cs:2.77555756156289e-17,0) rectangle (axis cs:0.33,81.2);
    \addlegendimage{ybar,ybar legend,draw=none,fill=chocolate2267451,very thin}
    \addlegendentry{Small-Scale Results}
        
    \draw[draw=none,fill=chocolate2267451,very thin] (axis cs:1,0) rectangle (axis cs:1.33,91);
    \draw[draw=none,fill=chocolate2267451,very thin] (axis cs:2,0) rectangle (axis cs:2.33,92.4);
    \draw[draw=none,fill=chocolate2267451,very thin] (axis cs:3,0) rectangle (axis cs:3.33,91.7);
    \draw[draw=none,fill=chocolate2267451,very thin] (axis cs:4,0) rectangle (axis cs:4.33,89.9);
    \draw[draw=none,fill=chocolate2267451,very thin] (axis cs:5,0) rectangle (axis cs:5.33,89.2);
    \draw[draw=none,fill=steelblue52138189,very thin] (axis cs:0.33,0) rectangle (axis cs:0.66,83.8);
    \addlegendimage{ybar,ybar legend,draw=none,fill=steelblue52138189,very thin}
    \addlegendentry{Model Predictions}
    
    \draw[draw=none,fill=steelblue52138189,very thin] (axis cs:1.33,0) rectangle (axis cs:1.66,88.9);
    \draw[draw=none,fill=steelblue52138189,very thin] (axis cs:2.33,0) rectangle (axis cs:2.66,95.6);
    \draw[draw=none,fill=steelblue52138189,very thin] (axis cs:3.33,0) rectangle (axis cs:3.66,92.9);
    \draw[draw=none,fill=steelblue52138189,very thin] (axis cs:4.33,0) rectangle (axis cs:4.66,87.9);
    \draw[draw=none,fill=steelblue52138189,very thin] (axis cs:5.33,0) rectangle (axis cs:5.66,89.8);
    \draw (axis cs:0.165,81.2) ++(0pt,2pt) node[
      scale=0.5,
      anchor=south,
      text=black,
      rotate=0.0
    ]{81.2};
    \draw (axis cs:1.165,91) ++(0pt,2pt) node[
      scale=0.5,
      anchor=south,
      text=black,
      rotate=0.0
    ]{91};
    \draw (axis cs:2.165,92.4) ++(0pt,2pt) node[
      scale=0.5,
      anchor=south,
      text=black,
      rotate=0.0
    ]{92.4};
    \draw (axis cs:3.165,91.7) ++(0pt,2pt) node[
      scale=0.5,
      anchor=south,
      text=black,
      rotate=0.0
    ]{91.7};
    \draw (axis cs:4.165,89.9) ++(0pt,2pt) node[
      scale=0.5,
      anchor=south,
      text=black,
      rotate=0.0
    ]{89.9};
    \draw (axis cs:5.165,89.2) ++(0pt,2pt) node[
      scale=0.5,
      anchor=south,
      text=black,
      rotate=0.0
    ]{89.2};
    \draw (axis cs:0.495,83.8) ++(0pt,2pt) node[
      scale=0.5,
      anchor=south,
      text=black,
      rotate=0.0
    ]{83.8};
    \draw (axis cs:1.495,88.9) ++(0pt,2pt) node[
      scale=0.5,
      anchor=south,
      text=black,
      rotate=0.0
    ]{88.9};
    \draw (axis cs:2.495,95.6) ++(0pt,2pt) node[
      scale=0.5,
      anchor=south,
      text=black,
      rotate=0.0
    ]{95.6};
    \draw (axis cs:3.495,92.9) ++(0pt,2pt) node[
      scale=0.5,
      anchor=south,
      text=black,
      rotate=0.0
    ]{92.9};
    \draw (axis cs:4.495,87.9) ++(0pt,2pt) node[
      scale=0.5,
      anchor=south,
      text=black,
      rotate=0.0
    ]{87.9};
    \draw (axis cs:5.495,89.8) ++(0pt,2pt) node[
      scale=0.5,
      anchor=south,
      text=black,
      rotate=0.0
    ]{89.8};
    \end{axis}
    
    \end{tikzpicture}
    \caption{A comparison of 8-subdomain runtimes.}
    \label{fig:runtime8}
\end{figure}
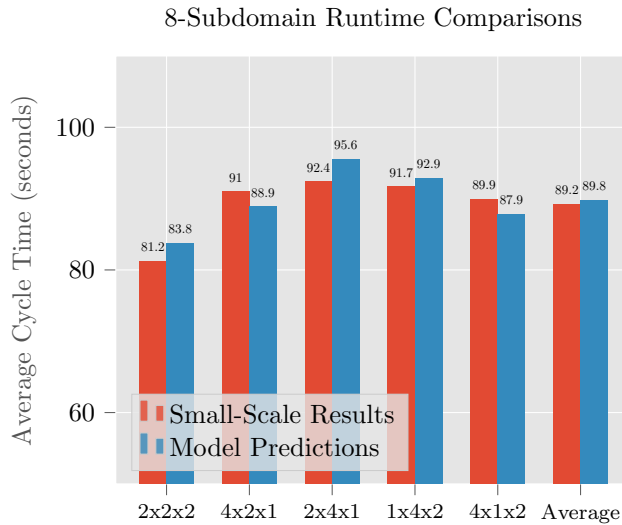

The computation load predictions from the model were used as inputs to the Shift processor allocation algorithm, and used to generate processor allocations for a set of large-scale simulations.
These simulations involved 10 inactive cycles, followed by 10 active cycles, with a starting particle count of 50 million.
The simulations were executed on 40 compute nodes of Lawrence Livermore National Lab's Lassen compute cluster, representing 800 MPI ranks.
Figure \ref{fig:runtime8} shows the average cycle times of the test simulations with a resolution of 8 subdomains; these have been selected for display as this resolution completed more quickly than others, allowing us to collect more test examples at this resolution.

At this resolution, the model's predictions produce simulations which complete at 99.3\% the speed of simulations using the small-scale results.
The first cycle, where memory is allocated and the problem tallies are initialized, takes much longer on average in simulations using both the small-scale results and the model's predictions, and have not been included in these comparisons to prevent them from dominating the average; however, all simulations have equivalent runtimes on these cycles as well.
The performance accuracy of the model is maintained across simulations with larger subdomains; with 16 subdomains, the model generates predictions that complete large-scale simulations at 98.2\% the speed of simulations using small-scale results; at 32 subdomains, simulations complete at 97.0\% speed.
This gives us an average runtime difference of 1.8\%, which falls well under the expected runtime of a small-scale simulation, which is roughly 10\% of the large-scale simulation time for a given problem.

\subsection{SMR Fuel Variants}
\label{ssec:variants}

One of the challenges discussed in section \ref{sec:motivation} was the training and data generation time needed to create an accurate model.
In order to justify the time needed to create a dataset and train the model on that dataset, the model must demonstrate some robustness to changes in the problems being simulated.
For the scale of this paper, only a single domain and problem type was generated for use as training data; for this reason, a good test of robustness would be to modify this domain and present modifications not seen in the original dataset.
This section shows the model's prediction results on two such modifications: a depleted fuel version of the SMR domain, and a version where the fuel assemblies were randomly shuffled within the SMR core.
No additional training was performed on the model to generate these predictions; the same training results generated in section \ref{ssec:ablation} were used for these simulations, and a matching representation was simply used as input to this version of the model.

\subsubsection{Depleted Fuel}
\label{ssec:depleted}

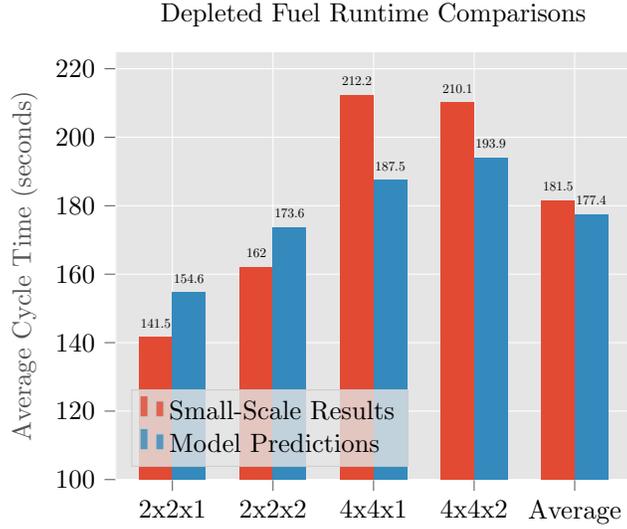
\begin{figure}[ht]
    \centering
    % This file was created with tikzplotlib v0.10.1.
    \begin{tikzpicture}
    
    \definecolor{chocolate2267451}{RGB}{226,74,51}
    \definecolor{dimgray85}{RGB}{85,85,85}
    \definecolor{gainsboro229}{RGB}{229,229,229}
    \definecolor{lightgray204}{RGB}{204,204,204}
    \definecolor{steelblue52138189}{RGB}{52,138,189}
    
    \begin{axis}[
    axis background/.style={fill=gainsboro229},
    axis line style={white},
    legend cell align={left},
    legend style={
      fill opacity=0.8,
      draw opacity=1,
      text opacity=1,
      at={(0.03,0.03)},
      anchor=south west,
      draw=lightgray204,
      fill=gainsboro229
    },
    tick align=outside,
    tick pos=left,
    title={Depleted Fuel Runtime Comparisons},
    x grid style={white},
    xmajorgrids,
    xmin=-0.233, xmax=4.893,
    xtick style={color=dimgray85},
    xtick={0.33,1.33,2.33,3.33,4.33},
    xticklabels={2x2x1,2x2x2,4x4x1,4x4x2,Average},
    y grid style={white},
    ylabel=\textcolor{dimgray85}{Average Cycle Time (seconds)},
    ymajorgrids,
    ymin=100, ymax=225,
    ytick style={color=dimgray85}
    ]
    \draw[draw=none,fill=chocolate2267451,very thin] (axis cs:2.77555756156289e-17,0) rectangle (axis cs:0.33,141.5);
    \addlegendimage{ybar,ybar legend,draw=none,fill=chocolate2267451,very thin}
    \addlegendentry{Small-Scale Results}
    
    \draw[draw=none,fill=chocolate2267451,very thin] (axis cs:1,0) rectangle (axis cs:1.33,162);
    \draw[draw=none,fill=chocolate2267451,very thin] (axis cs:2,0) rectangle (axis cs:2.33,212.2);
    \draw[draw=none,fill=chocolate2267451,very thin] (axis cs:3,0) rectangle (axis cs:3.33,210.1);
    \draw[draw=none,fill=chocolate2267451,very thin] (axis cs:4,0) rectangle (axis cs:4.33,181.5);
    \draw[draw=none,fill=steelblue52138189,very thin] (axis cs:0.33,0) rectangle (axis cs:0.66,154.6);
    \addlegendimage{ybar,ybar legend,draw=none,fill=steelblue52138189,very thin}
    \addlegendentry{Model Predictions}
    
    \draw[draw=none,fill=steelblue52138189,very thin] (axis cs:1.33,0) rectangle (axis cs:1.66,173.6);
    \draw[draw=none,fill=steelblue52138189,very thin] (axis cs:2.33,0) rectangle (axis cs:2.66,187.5);
    \draw[draw=none,fill=steelblue52138189,very thin] (axis cs:3.33,0) rectangle (axis cs:3.66,193.9);
    \draw[draw=none,fill=steelblue52138189,very thin] (axis cs:4.33,0) rectangle (axis cs:4.66,177.4);
    \draw (axis cs:0.165,141.5) ++(0pt,2pt) node[
      scale=0.5,
      anchor=south,
      text=black,
      rotate=0.0
    ]{141.5};
    \draw (axis cs:1.165,162) ++(0pt,2pt) node[
      scale=0.5,
      anchor=south,
      text=black,
      rotate=0.0
    ]{162};
    \draw (axis cs:2.165,212.2) ++(0pt,2pt) node[
      scale=0.5,
      anchor=south,
      text=black,
      rotate=0.0
    ]{212.2};
    \draw (axis cs:3.165,210.1) ++(0pt,2pt) node[
      scale=0.5,
      anchor=south,
      text=black,
      rotate=0.0
    ]{210.1};
    \draw (axis cs:4.165,181.5) ++(0pt,2pt) node[
      scale=0.5,
      anchor=south,
      text=black,
      rotate=0.0
    ]{181.5};
    \draw (axis cs:0.495,154.6) ++(0pt,2pt) node[
      scale=0.5,
      anchor=south,
      text=black,
      rotate=0.0
    ]{154.6};
    \draw (axis cs:1.495,173.6) ++(0pt,2pt) node[
      scale=0.5,
      anchor=south,
      text=black,
      rotate=0.0
    ]{173.6};
    \draw (axis cs:2.495,187.5) ++(0pt,2pt) node[
      scale=0.5,
      anchor=south,
      text=black,
      rotate=0.0
    ]{187.5};
    \draw (axis cs:3.495,193.9) ++(0pt,2pt) node[
      scale=0.5,
      anchor=south,
      text=black,
      rotate=0.0
    ]{193.9};
    \draw (axis cs:4.495,177.4) ++(0pt,2pt) node[
      scale=0.5,
      anchor=south,
      text=black,
      rotate=0.0
    ]{177.4};
    \end{axis}
    
    \end{tikzpicture}
    \caption{A comparison of depleted fuel SMR runtimes.}
    \label{fig:runtimedep}
\end{figure}

Figure \ref{fig:runtimedep} shows the results of a suite of large-scale depleted fuel SMR simulations, comparing the model's predictions to those generated by a small-scale simulation of the problem at several subdomain resolutions.
As the figure shows, the model's predictions perform as well as the small-scale predictions on this problem, achieving a slight performance increase of approximately 2.3\%.
This performance increase is most prominent in simulations with a higher number of subdomains; notably, the processor allocations generated by the small-scale predictions led to a crash in the large-scale simulations in both the 16- and 32-subdomain cases.
The model's predictions seemingly generated more stable processor allocations, which were able to complete a full large-scale simulation.
Due to the large amount of memory required to simulate the depleted fuel problem, it is possible that the current load prediction method is not sufficient for a domain of this complexity; this issue is discussed in greater detail in section \ref{sec:conclusion}.

\subsubsection{Shuffled Fuel}
\label{ssec:shuffled}

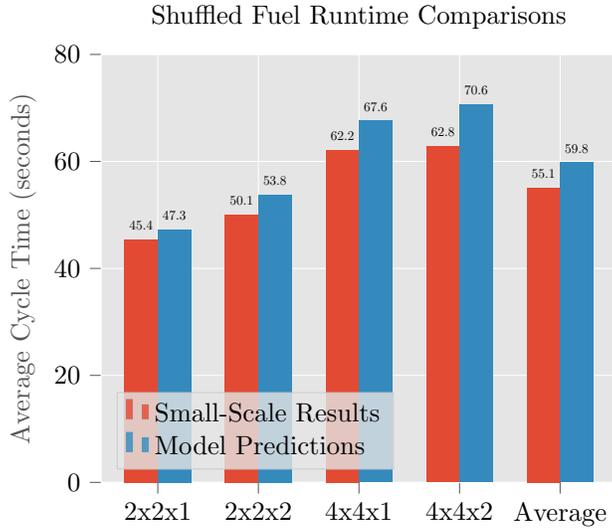
\begin{figure}[ht]
    \centering
        % This file was created with tikzplotlib v0.10.1.
    \begin{tikzpicture}
    
    \definecolor{chocolate2267451}{RGB}{226,74,51}
    \definecolor{dimgray85}{RGB}{85,85,85}
    \definecolor{gainsboro229}{RGB}{229,229,229}
    \definecolor{lightgray204}{RGB}{204,204,204}
    \definecolor{steelblue52138189}{RGB}{52,138,189}
    
    \begin{axis}[
    axis background/.style={fill=gainsboro229},
    axis line style={white},
    legend cell align={left},
    legend style={
      fill opacity=0.8,
      draw opacity=1,
      text opacity=1,
      at={(0.03,0.03)},
      anchor=south west,
      draw=lightgray204,
      fill=gainsboro229
    },
    tick align=outside,
    tick pos=left,
    title={Shuffled Fuel Runtime Comparisons},
    x grid style={white},
    xmajorgrids,
    xmin=-0.233, xmax=4.893,
    xtick style={color=dimgray85},
    xtick={0.33,1.33,2.33,3.33,4.33},
    xticklabels={2x2x1,2x2x2,4x4x1,4x4x2,Average},
    y grid style={white},
    ylabel=\textcolor{dimgray85}{Average Cycle Time (seconds)},
    ymajorgrids,
    ymin=0, ymax=80,
    ytick style={color=dimgray85}
    ]
    \draw[draw=none,fill=chocolate2267451,very thin] (axis cs:2.77555756156289e-17,0) rectangle (axis cs:0.33,45.4);
    \addlegendimage{ybar,ybar legend,draw=none,fill=chocolate2267451,very thin}
    \addlegendentry{Small-Scale Results}
    
    \draw[draw=none,fill=chocolate2267451,very thin] (axis cs:1,0) rectangle (axis cs:1.33,50.1);
    \draw[draw=none,fill=chocolate2267451,very thin] (axis cs:2,0) rectangle (axis cs:2.33,62.2);
    \draw[draw=none,fill=chocolate2267451,very thin] (axis cs:3,0) rectangle (axis cs:3.33,62.8);
    \draw[draw=none,fill=chocolate2267451,very thin] (axis cs:4,0) rectangle (axis cs:4.33,55.1);
    \draw[draw=none,fill=steelblue52138189,very thin] (axis cs:0.33,0) rectangle (axis cs:0.66,47.3);
    \addlegendimage{ybar,ybar legend,draw=none,fill=steelblue52138189,very thin}
    \addlegendentry{Model Predictions}
    
    \draw[draw=none,fill=steelblue52138189,very thin] (axis cs:1.33,0) rectangle (axis cs:1.66,53.8);
    \draw[draw=none,fill=steelblue52138189,very thin] (axis cs:2.33,0) rectangle (axis cs:2.66,67.6);
    \draw[draw=none,fill=steelblue52138189,very thin] (axis cs:3.33,0) rectangle (axis cs:3.66,70.6);
    \draw[draw=none,fill=steelblue52138189,very thin] (axis cs:4.33,0) rectangle (axis cs:4.66,59.8);
    \draw (axis cs:0.165,45.4) ++(0pt,2pt) node[
      scale=0.5,
      anchor=south,
      text=black,
      rotate=0.0
    ]{45.4};
    \draw (axis cs:1.165,50.1) ++(0pt,2pt) node[
      scale=0.5,
      anchor=south,
      text=black,
      rotate=0.0
    ]{50.1};
    \draw (axis cs:2.165,62.2) ++(0pt,2pt) node[
      scale=0.5,
      anchor=south,
      text=black,
      rotate=0.0
    ]{62.2};
    \draw (axis cs:3.165,62.8) ++(0pt,2pt) node[
      scale=0.5,
      anchor=south,
      text=black,
      rotate=0.0
    ]{62.8};
    \draw (axis cs:4.165,55.1) ++(0pt,2pt) node[
      scale=0.5,
      anchor=south,
      text=black,
      rotate=0.0
    ]{55.1};
    \draw (axis cs:0.495,47.3) ++(0pt,2pt) node[
      scale=0.5,
      anchor=south,
      text=black,
      rotate=0.0
    ]{47.3};
    \draw (axis cs:1.495,53.8) ++(0pt,2pt) node[
      scale=0.5,
      anchor=south,
      text=black,
      rotate=0.0
    ]{53.8};
    \draw (axis cs:2.495,67.6) ++(0pt,2pt) node[
      scale=0.5,
      anchor=south,
      text=black,
      rotate=0.0
    ]{67.6};
    \draw (axis cs:3.495,70.6) ++(0pt,2pt) node[
      scale=0.5,
      anchor=south,
      text=black,
      rotate=0.0
    ]{70.6};
    \draw (axis cs:4.495,59.8) ++(0pt,2pt) node[
      scale=0.5,
      anchor=south,
      text=black,
      rotate=0.0
    ]{59.8};
    \end{axis}
    
    \end{tikzpicture}
    \caption{A comparison of shuffled fuel SMR runtimes.}
    \label{fig:runtimeshuf}
\end{figure}

A suite of shuffled fuel SMR domains were created by randomly assigning one of the three fuel assembly types found in the fresh-fuel SMR domain to each assembly space in the SMR core architecture.
To ensure that this random assignment did not introduce a bias for either methodology, the average runtime of three such assignments was taken for each method; Figure \ref{fig:runtimeshuf} shows this average runtime for each method at several subdomain resolutions.
As the figure shows, the model's predictions are approximately 8.5\% slower than the small-scale predictions on average; this is still within the 10\% runtime taken to generate a small-scale prediction, but does seem to represent the limits of what this version of the model can reliably predict.
A brief re-training run using a small dataset of shuffled SMR domains may allow the model to more accurately learn the performance of these configurations, and improve results on these shuffled fuel simulations.

\subsection{C5G7 Results}
\label{ssec:c5g7}

%\begin{wrapfigure}{r}{0.5\textwidth}
%    \centering
%        \includegraphics[width=0.48\textwidth]{c5g7_idx.png}
%    \caption{A z-axial slice of the C5G7 domain, and the indexing used for dataset generation.}
%    \label{fig:c5g7}
%\end{wrapfigure}

\begin{figure}[h]
    \centering
        \includegraphics[width=0.48\textwidth]{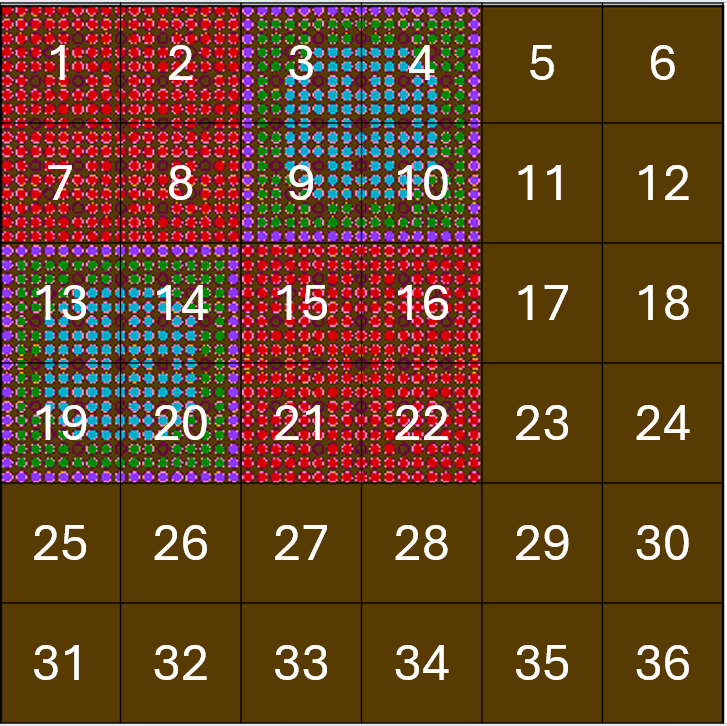}
    \caption{A z-axial slice of the C5G7 domain, and the indexing used for dataset generation.}
    \label{fig:c5g7}
\end{figure}

In order to show the model's performance on other types of problem geometries, a new dataset was created for the C5G7 domain.
This dataset was generated using a similar method to the one described in Section \ref{ssec:dataset}; however, due to the structure of the C5G7 domain (seen in Figure \ref{fig:c5g7}), the domain was instead divided into a 6x6x6 grid of elementary volumes.
The model architecture described in Section \ref{ssec:architecture} was similarly modified to account for this new input size.

\subsubsection{Model Performance}
\label{ssec:c5g7performance}

\begin{figure}[ht]
    \centering
    \input{model_c5g7}
    \caption{A graph of the model's training and validation losses during a training run on the C5G7 dataset.}
    \label{fig:modelc5g7}
\end{figure}

Figure \ref{fig:modelc5g7} shows the training and validation losses of the model's training session over 4000 epochs on four AMD MI-250X GPUs; this training run lasted approximately five hours on this hardware.
The model used the same learning rate, step factor, and loss equation derived in section \ref{ssec:ablation}.
The model achieved a minimum validation loss of 10.35; this loss value is notably larger than the minimum loss seen in section \ref{ssec:performance}.
This may be due to the fact that the range and average size of computation load values were both larger than those seen in the SMR dataset; including more nuclide data in the input embedding of the model's architecture may improve accuracy on this problem.
Across the entire validation set, the model achieves an $R^2$ score of $98.0\%$.
The model was trained on 2000 datapoints; each point represents an input/output pair of a C5G7 domain decomposition and its corresponding computation load, as described in section \ref{ssec:dataset}.

\subsubsection{Simulation Runtimes}
\label{ssec:c5g7runtimes}

\begin{figure}[ht]
    \centering
        % This file was created with tikzplotlib v0.10.1.
    \begin{tikzpicture}
    
    \definecolor{chocolate2267451}{RGB}{226,74,51}
    \definecolor{dimgray85}{RGB}{85,85,85}
    \definecolor{gainsboro229}{RGB}{229,229,229}
    \definecolor{lightgray204}{RGB}{204,204,204}
    \definecolor{steelblue52138189}{RGB}{52,138,189}
    
    \begin{axis}[
    axis background/.style={fill=gainsboro229},
    axis line style={white},
    legend cell align={left},
    legend style={
      fill opacity=0.8,
      draw opacity=1,
      text opacity=1,
      at={(0.03,0.03)},
      anchor=south west,
      draw=lightgray204,
      fill=gainsboro229
    },
    tick align=outside,
    tick pos=left,
    title={8-Subdomain Runtime Comparisons - C5G7},
    x grid style={white},
    xmajorgrids,
    xmin=-0.283, xmax=5.943,
    xtick style={color=dimgray85},
    xtick={0.33,1.33,2.33,3.33,4.33,5.33},
    xticklabels={2x2x2,4x2x1,2x4x1,1x4x2,4x1x2,Average},
    y grid style={white},
    ylabel=\textcolor{dimgray85}{Average Cycle Time (seconds)},
    ymajorgrids,
    ymin=50, ymax=85,
    ytick style={color=dimgray85}
    ]
    \draw[draw=none,fill=chocolate2267451,very thin] (axis cs:2.77555756156289e-17,0) rectangle (axis cs:0.33,78.5);
    \addlegendimage{ybar,ybar legend,draw=none,fill=chocolate2267451,very thin}
    \addlegendentry{Small-Scale Results}
    
    \draw[draw=none,fill=chocolate2267451,very thin] (axis cs:1,0) rectangle (axis cs:1.33,65.7);
    \draw[draw=none,fill=chocolate2267451,very thin] (axis cs:2,0) rectangle (axis cs:2.33,66.7);
    \draw[draw=none,fill=chocolate2267451,very thin] (axis cs:3,0) rectangle (axis cs:3.33,66.6);
    \draw[draw=none,fill=chocolate2267451,very thin] (axis cs:4,0) rectangle (axis cs:4.33,66.4);
    \draw[draw=none,fill=chocolate2267451,very thin] (axis cs:5,0) rectangle (axis cs:5.33,68.8);
    \draw[draw=none,fill=steelblue52138189,very thin] (axis cs:0.33,0) rectangle (axis cs:0.66,80);
    \addlegendimage{ybar,ybar legend,draw=none,fill=steelblue52138189,very thin}
    \addlegendentry{Model Predictions}
    
    \draw[draw=none,fill=steelblue52138189,very thin] (axis cs:1.33,0) rectangle (axis cs:1.66,62.7);
    \draw[draw=none,fill=steelblue52138189,very thin] (axis cs:2.33,0) rectangle (axis cs:2.66,68.2);
    \draw[draw=none,fill=steelblue52138189,very thin] (axis cs:3.33,0) rectangle (axis cs:3.66,67.9);
    \draw[draw=none,fill=steelblue52138189,very thin] (axis cs:4.33,0) rectangle (axis cs:4.66,67.8);
    \draw[draw=none,fill=steelblue52138189,very thin] (axis cs:5.33,0) rectangle (axis cs:5.66,69.3);
    \draw (axis cs:0.165,78.5) ++(0pt,2pt) node[
      scale=0.5,
      anchor=south,
      text=black,
      rotate=0.0
    ]{78.5};
    \draw (axis cs:1.165,65.7) ++(0pt,2pt) node[
      scale=0.5,
      anchor=south,
      text=black,
      rotate=0.0
    ]{65.7};
    \draw (axis cs:2.165,66.7) ++(0pt,2pt) node[
      scale=0.5,
      anchor=south,
      text=black,
      rotate=0.0
    ]{66.7};
    \draw (axis cs:3.165,66.6) ++(0pt,2pt) node[
      scale=0.5,
      anchor=south,
      text=black,
      rotate=0.0
    ]{66.6};
    \draw (axis cs:4.165,66.4) ++(0pt,2pt) node[
      scale=0.5,
      anchor=south,
      text=black,
      rotate=0.0
    ]{66.4};
    \draw (axis cs:5.165,68.8) ++(0pt,2pt) node[
      scale=0.5,
      anchor=south,
      text=black,
      rotate=0.0
    ]{68.8};
    \draw (axis cs:0.495,80) ++(0pt,2pt) node[
      scale=0.5,
      anchor=south,
      text=black,
      rotate=0.0
    ]{80};
    \draw (axis cs:1.495,62.7) ++(0pt,2pt) node[
      scale=0.5,
      anchor=south,
      text=black,
      rotate=0.0
    ]{62.7};
    \draw (axis cs:2.495,68.2) ++(0pt,2pt) node[
      scale=0.5,
      anchor=south,
      text=black,
      rotate=0.0
    ]{68.2};
    \draw (axis cs:3.495,67.9) ++(0pt,2pt) node[
      scale=0.5,
      anchor=south,
      text=black,
      rotate=0.0
    ]{67.9};
    \draw (axis cs:4.495,67.8) ++(0pt,2pt) node[
      scale=0.5,
      anchor=south,
      text=black,
      rotate=0.0
    ]{67.8};
    \draw (axis cs:5.495,69.3) ++(0pt,2pt) node[
      scale=0.5,
      anchor=south,
      text=black,
      rotate=0.0
    ]{69.3};
    \end{axis}
    
    \end{tikzpicture}
    \caption{A comparison of C5G7 domain runtimes.}
    \label{fig:c5g7runtime8}
\end{figure}
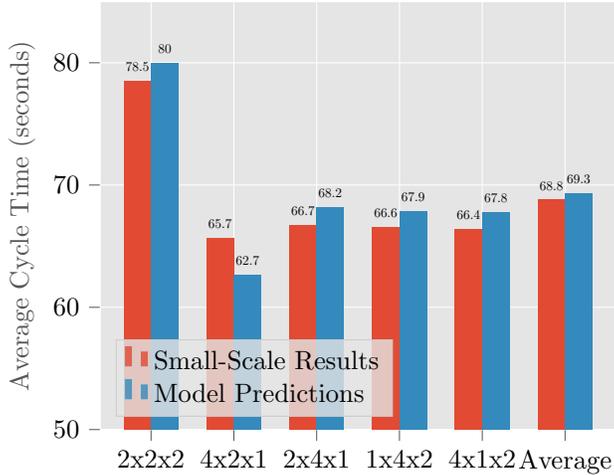

Similarly to the results seen in section \ref{ssec:runtimes}, these results focus on the large-scale cycle runtimes of a suite of 8, 16, and 32-subdomain C5G7 simulations, using 10 inactive cycles followed by 10 active cycles, with an initial particle count of 50 million, executed on 40 compute nodes on Lassen (800 MPI ranks).
Figure \ref{fig:c5g7runtime8} shows the average cycle times of the test simulations with a resolution of 8 subdomains.
The model's predictions generated processor allocations that completed at 99.3\% the speed of simulations using the small-scale results.
Much like the SMR simulations, we see similar results across all subdomain resolutions; in the 16-subdomain suite, the model's predictions perform at 97.6\% speed compared to the small-scale results, and in the 32-subdomain suite, simulations complete at 98.5\% speed.
This gives us an average runtime difference of 1.5\%; this change in runtime is similar to the change seen in the SMR domain, indicating that the model's performance does not seem to significantly change when trained on a different reactor geometry.

\subsection{Parameter Sensitivity}
\label{ssec:sensitivity}

Our final study of the model's capabilities was to examine its sensitivity to changes in simulation parameters; in other words, to see if it has a robustness to these changes that is not present in the current methodology, as discussed in section \ref{sec:motivation}.
It should be noted that this robustness represents a major challenge, due to the wider variance in computation load created by modifying the parameters of these simulations.
In order to test this robustness, we added a set of small-scale fresh-fuel SMR simulations to the dataset which contained a varying number of processors, cycles, and particle histories.
This dataset was used to re-train the original model, using its pre-existing weights as a starting point for fine-tuning.
As this re-training was performed, it became clear that the model needed to be given the simulation parameters as part of its input in order to be able to make accurate predictions.
This led to a new challenge: adjusting the model architecture in a way that retains the training that has been done, while incorporating the simulation parameters into the input of the model itself.

\begin{figure}[ht]
    \centering
    \input{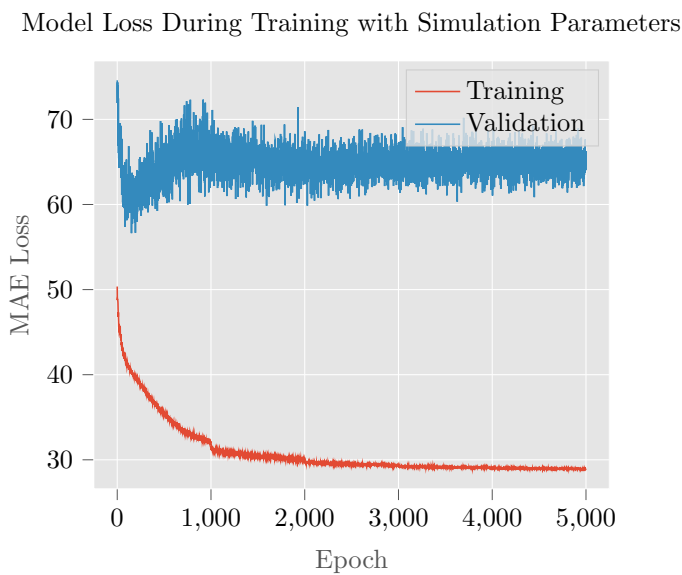}
    \caption{A graph of the model's training and validation losses during a training run with variation in simulation parameters.}
    \label{fig:paramloss}
\end{figure}

Several methods to achieve this goal were attempted; the method with the highest accuracy will be highlighted in this section.
Rather than modifying the Transformer architecture, the most successful method found in our preliminary search was to add a multi-layer perceptron (MLP) model to the output side of the architecture.
In other words, an MLP was added to the flow of the model which accepts the output of the Transformer and the parameter values as its own input, and produces a new computation load prediction as its output.
This MLP consists of two linear layers, and its weights are trained independently of the Transformer's weights; no additional training was performed on the Transformer model, instead using the pre-trained weights from Section \ref{ssec:performance}.

Figure \ref{fig:paramloss} shows the MAE loss values of a training run using this model architecture.
Due to the increased size of the dataset and simpler architecture, the MLP was trained over 5000 epochs.
As we can see, the model begins its training run with a similar loss curve to the Transformer training regimens; however, its loss values level out around 250 epochs.
The lowest MAE loss value achieved by the model during training is 56.6; this is larger than the loss values from the single parameter dataset seen in Section \ref{ssec:performance}, which is to be expected due to the increased variance in outputs in the multi-parameter dataset.
Further work is required to create a model that can generate accurate predictions when trained on simulations with a variety of parameters; however, this method of fine-tuning predictions using a secondary model seems promising, and more complex model architectures may lead to improved results.

\section{Conclusion}
\label{sec:conclusion}

In this paper, we sought to train a Transformer model on domain-decomposed SMR simulations to generate processor allocations without the need to run a small-scale simulation; specifically, our goal was to develop a model that can produce results with equivalent accuracy to the current methodology in less time and with less computational overhead.
In that regard, this model was largely successful; large-scale simulations using the predictions generated by the model ran at roughly 98\% the speed of simulations using state-of-the-art load prediction methods, while skipping the small-scale simulation step entirely.
These predictions were generated more quickly than the current methodology, providing an estimation of computational load with similar accuracy at half of the overall runtime, even when accounting for time taken to initialize the model and load its trained weights.
This speed-up also occurs despite running on only 4 GPUs, compared to the 800 CPUs used to simulate small-scale predictions.
Multiple modifications to the domain's fuel assemblies were performed to test the model's robustness to data not present within its training set; these modifications did not heavily impact the model's performance.
A low number of particle histories were used in the simulations that comprise the model's dataset, to more evenly compare the model's predictions to those generated by the current methodology; however, the model's accuracy on large-scale simulations could likely be improved by training on simulations with a higher number of particle histories.
Currently, the model does not appear to be robust enough to predict accurately on simulations with a variety of parameters; after a preliminary study, it was determined that more work will be needed in the future to improve this robustness.

Further developments could include the creation of a ``domain-agnostic" version of the model, which can accept any LWR geometry generated by ORNL's VERA tool.
This version of the model would be closer to the ideal model architecture described in section \ref{sec:motivation}, and would require a great deal more data and training time.
Once such a model is successfully created, a method of selecting an optimal domain decomposition from a given problem input naturally follows, given this model will have the necessary data to make that decision for an end user without the need for ``guess-and-check".
Additionally, there are a number of Monte Carlo particle transport codes that simulate problems and domains similar to those seen in Shift; this model could easily be transferred to other code bases and re-trained for their specific applications.
The model may also be able to generate predictions on other particle transport problems, such as fixed-source or charged particle simulations, without significant additional work.

Lastly, the processor allocation algorithm used in this project may merit re-evaluation after some insights gained from testing this model.
During testing, it became clear that small-load subdomains represented a special challenge that may not be addressed by the current methodology.
These subdomains, although small in computational load, still contain a significant memory overhead that caused our simulations to crash or take far longer to complete when too few processors (and thus too little memory) were allocated; this was noted on both the model's predicted load values, and those obtained from small-scale simulation results.
The sensitivity of these subdomains, and their impact on simulation stability and runtime, require further analysis to understand and improve processor allocation on these types of problems.

\section*{Acknowledgment}

The authors would like to thank Aaron Reynolds for his help with the Shift frontend used to generate this dataset and compare results, Steven Hamilton and Tom Evans for providing the input decks used to run simulations on Shift, and Austin Ellis for providing insights into the processor allocation algorithm used in our results.

\vspace{0.1in}
This work was supported by the Center for Exascale Monte-Carlo Neutron Transport (CEMeNT), a PSAAP-III project funded by the Department of Energy, grant number: DE-NA003967. The authors report there are no competing interests to declare.

\pagebreak
\bibliographystyle{style/tfcad}
\bibliography{bibliography}

\end{document}